\documentclass[%
 aip,
rsi,%
 amsmath,amssymb,
reprint,%
]{revtex4-1}

\newcommand{\tj}[6]{ \begin{pmatrix}
       #1 & #2 & #3 \\
       #4 & #5 & #6 
\end{pmatrix}}

\def\bnabla{\mbox{\boldmath $\nabla $}}

\newcommand{\ee}{\overline{\varepsilon}}
\usepackage{comment}

\usepackage{graphicx}
\usepackage{dcolumn}
\usepackage{bm}
\usepackage{comment}
\usepackage{color}
\usepackage{tcolorbox}
\usepackage{tabularx}
\usepackage{array}
\usepackage{colortbl}
\tcbuselibrary{skins}
\usepackage{hyperref}

\makeatletter
\newcommand*{\textoverline}[1]{$\overline{\hbox{#1}}\m@th$}
\makeatother

\begin{document}


\title[Charge regulation of patchy charged particles]{Anomalous Multipole Expansion: Charge Regulation of Patchy, Inhomogeneously Charged Spherical Particles}

\author{An\v ze \surname{Lo\v sdorfer Bo\v zi\v c}}
\email{anze.bozic@ijs.si}
\affiliation{Department of Theoretical Physics, Jo\v zef Stefan Institute, SI-1000 Ljubljana, Slovenia}
\author{Rudolf Podgornik}
\affiliation{Department of Theoretical Physics, Jo\v zef Stefan Institute, SI-1000 Ljubljana, Slovenia}
\affiliation{Department of Physics, Faculty of Mathematics and Physics, University of Ljubljana, SI-1000 Ljubljana, Slovenia}
\affiliation{School of Physical Sciences and Kavli Institute for Theoretical Sciences, University of Chinese Academy of Sciences, Beijing 100049, China} 
\affiliation{CAS Key Laboratory of Soft Matter Physics, Institute of Physics, Chinese Academy of Sciences, Beijing 100190, China}

\date{\today}

\begin{abstract}
Charge regulation is an important aspect of electrostatics in biological and colloidal systems, where the charges are generally not fixed, but depend on the environmental variables. Here, we analyze the charge regulation mechanism in patchy, inhomogeneously charged spherical particles, such as globular proteins, colloids, or viruses. Together with the multipole expansion of inhomogeneously charged spherical surfaces, the charge regulation mechanism on the level of linear approximation is shown to lead to a mixing between different multipole moments depending on their capacitance -- the response function of the charge distribution with respect to the electrostatic potential. This presents an additional anomalous feature of molecular electrostatics in the presence of ionic screening. We demonstrate the influence of charge regulation on several examples of inhomogeneously charged spherical particles, showing that it leads to significant changes in their multipole moments.
\end{abstract}

\maketitle

\section{\label{sec:intro}Introduction}

In general, charges in biological and colloidal systems are not fixed, but result from some charge separation mechanism stemming from dissociation/association equilibria. Typical examples are the acid/base equilibria of dissolved salts and minerals, dissociation of protonating/deprotonating titrating groups in membrane lipids or protein amino acids, and adsorption of solution ions onto solid substrates in aqueous solutions~\cite{Borkovec2001}. These charge separation mechanisms all involve a specific chemical component, which can be described both by a chemical equilibrium constant, arising from detailed quantum chemical considerations, as well as by a more coarse-grained electrostatic component, originating in the dissociated charges. The latter component was first considered in the context of ``protein ionization'' by Linderstr{\o}m-Lang in the 1920s~\cite{Lang1924}. The conceptual basis of these acid/base equilibria was developed by Marcus~\cite{Marcus1955}, who considered the effect of solution electrostatic interactions on the fractional charging and titration curves of general polyelectrolytes in aqueous electrolyte solutions as a function of the ionic strength of the bathing solution. A similar approach for the description of protein titration curves within a dielectric continuum model was developed by Tanford and Kirkwood~\cite{Tanford1957a,Tanford1957b}. The framework of charge separation mechanisms was substantially widened when Ninham and Parsegian~\cite{Ninham1971} coupled the charge dissociation processes with the interactions between dissociable surfaces within the paradigm of {\em charge regulation} (CR), which found application in various domains of the theory of the stability of colloids~\cite{Chan1975,Prieve1976}. In more recent decades, CR was formalized for surface binding either by using the law of mass action~\cite{PericetCamara2004,Grunberg1999}, or equivalently, by using a model-specific surface free energy~\cite{Podgornik1995,Henle2004,Longo2012,Longo2013,Adzic2014,Adzic2015,Maggs2014,Markovich2014,Markovich2016,Diamant1996,Biesheuvel2004,BenYaakov2011}. CR is also closely related to the binding of ions to small molecules, proteins, polymers, colloid particles, and membranes, as well as to proton binding to these substrates, and has been reviewed in detail~\cite{Borkovec2001,Trefalt2016}.

Electrostatic interactions dominate many aspects of protein behavior, and cannot be properly understood without paying due attention to the protonation and deprotonation of their constituent ionizable amino acid residues~\cite{Warshel2006,Gitlin2006}, with the implied CR of the protein-specific distribution of dissociable charges~\cite{Lund2005,Krishnan2017}. In order to quantify the electrostatic interactions in proteins, one needs to encode not only the magnitudes of their charges, but also the anisotropic part of their distribution along their solvent-exposed molecular surface~\cite{Gramada2006,Hoppe2013}. This naturally leads to a multipole expansion of the protein surface charge density~\cite{ALB2017a}. Multipole moments of high order are known to influence the phase equilibria of concentrated protein solutions~\cite{Bianchi2017}, orientationally steer protein complexes into place~\cite{Abrikosov2017}, and can be used as a versatile phenomenological tool to dress up bare spheres in the first step towards the consideration of more complicated details of both protein charge distributions as well as patchy charge distributions in general~\cite{Boon2011,Lund2016}. Each multipole in this expansion series describes a particular charge motif, starting from the simplest monopolar, spherically-symmetric distribution~\cite{Paulini2005}. The most straightforward representation of the multipole expansion is obtained by mapping the charge distribution on the original solvent-accessible protein surface onto a sphere circumscribed to the protein~\cite{Postarnakevich2009,Arzensek2015,ALB2017a,ALB2018a}. Such a multipole expansion provides a mapping between a coarse-grained collective description of the charge density and the underlying detailed microscopic charge site distribution, so that any level of detail can be reached if enough multipole orders of the expansion are taken into account.

While a small number of multipole moments can often be used as a proxy to characterize the microscopic details of charge distributions in charged macromolecules~\cite{Gramada2006,Hoppe2013}, the details of the standard (Coulomb) multipole expansion~\cite{Schwinger} are substantially modified when the macromolecules are placed in a screening environment of aqueous bathing solutions. In fact, contrary to the standard multipole expansion, the screened electrostatic potential retains the full directional dependence for all multipole moments, so that the effects of charge anisotropy and high-order multipole moments extend all the way to the far-field region~\cite{Rowan2000,Kjellander2008,ALB2013a,Kjellander2016}. This entails an admixture of higher multipole moments to all the low-order multipoles -- including the monopole -- so that the usual argument that at large separations between charge distributions only the monopole moment matters is not substantiated. While this anomalous property of screened multipole electrostatics has been derived some time ago, it is often overlooked, e.g., in the modelling of protein-protein interactions.

In what follows, we will show that the CR mechanism in conjunction with the multipole moment expansion of inhomogeneously charged spherical surfaces leads to an additional anomalous feature of molecular electrostatics, similar to the full directional dependence of all multipole moments in a screened environment, yet differing in relevant details. CR mechanism will be shown to imply a mixing between different multipoles depending on their capacitance, i.e., the response function of the charge distribution with respect to the electrostatic potential. We will base our approach on a reformulation of the CR problem, akin to the framework set forth by Marcus~\cite{Marcus1955}: We will start with the multipole expansion of the Debye-H\"{u}ckel (DH) electrostatic energy for general surface charge distributions, weaving them afterwards into the CR theory through the corresponding CR free energy terms. The nature of the CR of dissociable groups will immediately yield an equation coupling a given multipole moment to other multipole moments. After deriving a linearized form of the CR theory, fully consistent with the DH approximation, we will provide a comprehensive discussion of these anomalous CR effects on the electrostatic multipoles of patchy spherical colloids and globular proteins with inhomogeneous surface charge distributions.

\section{Spherical surface charge density in presence of charge regulation}

\subsection{Model of surface charge density}

\begin{figure*}[!th]
\begin{center}
\includegraphics[width=1.75\columnwidth]{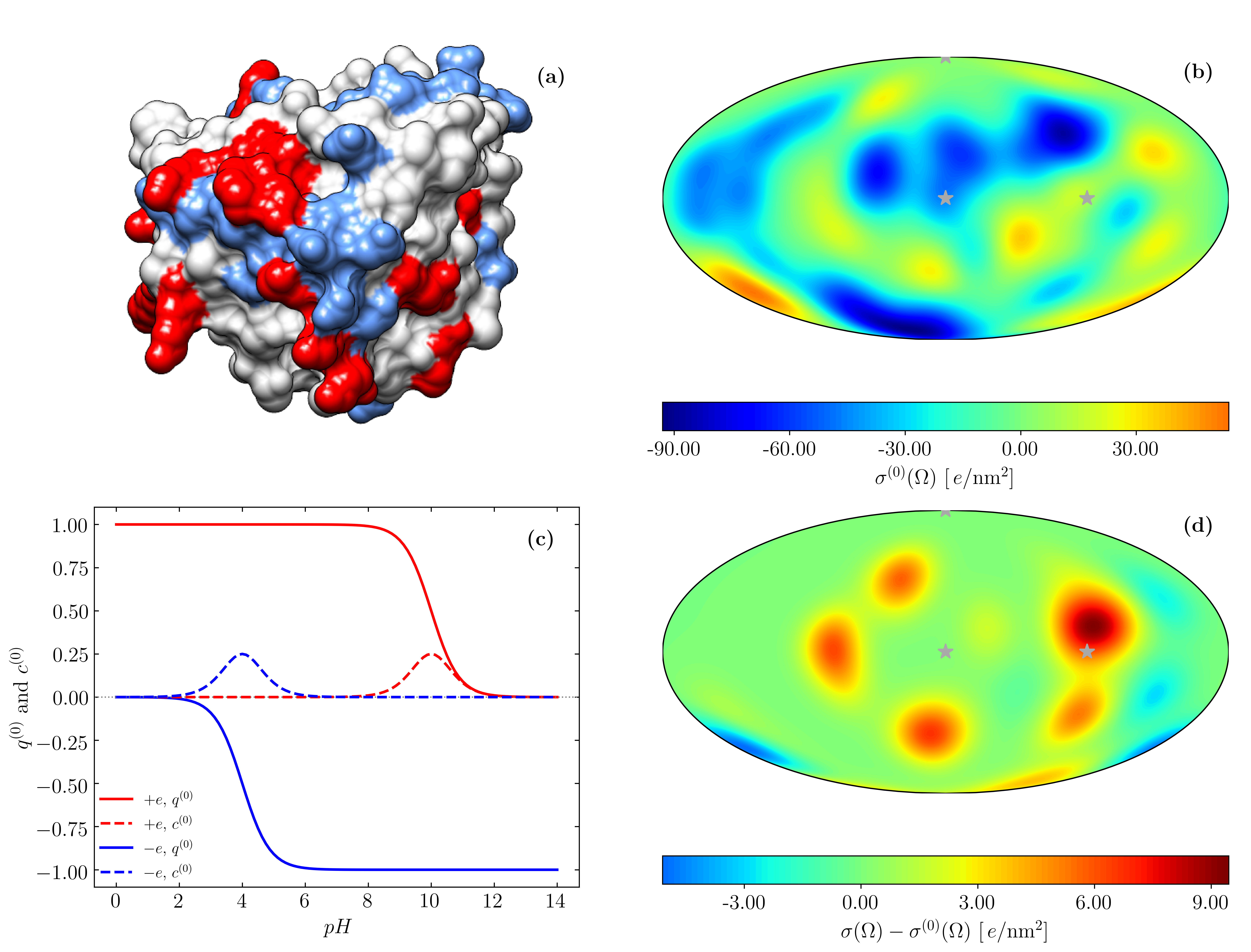}
\end{center}
\caption{{\bf (a)} An example of a globular protein, $\beta$-lactoglobulin (PDB: 2BLG), consisting of $52$ charged amino acids with different $pK_a$ values. Coloured in red are the amino acids which can become positively charged, and in blue the amino acids which can become negatively charged. The charge distribution of the protein can be mapped onto its circumscribed sphere with radius $R=1.06$ nm~\cite{ALB2017a}. Panel {\bf (b)} shows the ``bare'' surface charge distribution (in absence of CR) of 2BLG projected onto a plane using the Mollweide projection~\cite{ALB2013a,ALB2017a}. The distribution is shown at $pH=10$, $n_0=100$ mM, $\varepsilon_p=4$, and $\lambda=20$; these parameters are described in more detail in the main text. Grey stars show the positions of the Cartesian coordinate axes. {\bf (c)} Linearized CR mechanism, shown on the examples of a positive charge with $pK_a=10$ and a negative charge with $pK_a=3$. Linearized CR contribution to each charge consists of its ($pH$-dependent) bare charge $q^{(0)}$ and bare capacitance $c^{(0)}$ [Eqs.~\eqref{eq:barechr} and ~\eqref{eq:barecap}]. This mechanism leads to deviations from the bare charge distribution in absence of CR. Panel {\bf (d)} shows the difference between the surface charge distributions of 2BLG at $pH=10$ in presence and absence of CR. Other parameters of the distributions are the same as in panel {\bf (b)}.
\label{fig:1}}
\end{figure*}

While our derivation of the CR mechanism can be generalized to arbitrary inhomogeneously charged surfaces (with~\cite{Boon2011} or without~\cite{ALB2017a} any symmetry), our main concern will be spherical particles, such as patchy colloids and globular proteins (Fig.~\ref{fig:1}a). To model their surface charge densities, we will assume that all association/dissociation sites $\eta_k$ are located on a sphere with radius $R$. The charge on each of these sites, $q_k$, will be modelled by a normal distribution on the sphere, characterized by its mean direction $\Omega_k=(\vartheta_k,\varphi_k)$ and the spread around this direction $\lambda_k$ (the ``patchiness'' of the charge)~\cite{ALB2018b}. In addition, the surface charge distribution can consist of two different types of moieties: The first one can acquire a positive charge by protonation, and we write $q_k^+=e\eta_k$, where $\eta_k\in[0,1]$. The second one acquires a negative charge by deprotonation, and we write $q_k^-=e(\eta_k-1)$, where again $\eta_k\in[0,1]$.

The total surface charge distribution on the sphere is then given by~\cite{ALB2018b}
\begin{equation}
\label{eq:sig}
\sigma(\Omega)=\frac{1}{4\pi R^2}\sum_{k=1}^Nq_k^\pm\,\frac{\lambda_k}{\sinh\lambda_k}\exp(\lambda_k\cos\gamma_k),
\end{equation}
where $\cos\gamma_k$ is the great circle distance between $\Omega$ and $\Omega_k$. When $\lambda\to0$, the distribution becomes uniform on the sphere, whereas in the opposite limit of $\lambda\to\infty$, we obtain a distribution composed of point charges described by Dirac delta functions. Details of this model are given in Ref.~\onlinecite{ALB2018b}. Figure~\ref{fig:1}b shows an example of the surface charge distribution of a globular protein with $\lambda=20$ for all the constituent charges, mapped from a sphere onto a plane using the Mollweide projection~\cite{ALB2013a,ALB2017a}.

Writing the contribution of an individual charge as $\sigma^{(k)}(\Omega)$, we expand it in terms of multipole moments as
\begin{equation}
\label{eq:mexp}
\sigma^{(k)}(\Omega)=\frac{1}{4\pi R^2}\sum_{l,m}\sigma^{(k)}(lm)\,Y_{lm}(\Omega).
\end{equation}
The corresponding multipole expansion coefficients are~\cite{ALB2018b}
\begin{equation}
\label{eq:slm}
\sigma^{(k)}(lm)=4\pi\,q_k\,g_l(\lambda_k)\,Y_{lm}^*(\Omega_k),
\end{equation}
where we have introduced the function
\begin{equation}
\label{eq:gl}
g_l(\lambda)=\frac{\lambda}{\sinh\lambda}\,i_l(\lambda),
\end{equation}
and $i_l(x)$ are the modified spherical Bessel functions of the first kind. The multipole coefficients of the total surface charge density are then simply the sum of the contributions of the coefficients of individual charges, $\sigma(lm)=\sum_k\sigma^{(k)}(lm)$.

\subsection{Charge regulation}

The surface charge density on the spheres should not, however, be assumed a priori, but should follow from the minimization of the relevant total thermodynamic potential, yielding the equilibrium state in terms of equilibrium charge densities on the surface without any additional assumptions. This CR mechanism can be formalized either by invoking the chemical dissociation equilibrium of the surface binding sites with the corresponding law of mass action, or equivalently by adding a model surface free energy to the Poisson-Boltzmann (PB) or DH bulk free energy. The latter approach then leads to the same basic self-consistent CR boundary conditions for surface dissociation equilibrium through a minimization procedure, but without an explicit connection with the law of mass action. We will use this approach which will allow us to derive some useful approximations.

We first write down the total free energy of the system:
\begin{equation}
\label{eq:fe}
F=F_{DH}[\psi]+F_{CR}[\eta_k].
\end{equation}
The first term, $F_{DH}[\psi]$, is the DH free energy, dependent on the electrostatic potential $\psi$ of the system arising from the charge density $\sigma(\Omega)$ on the surface of the sphere. The DH free energy has the standard form
\begin{eqnarray}
\label{eq:fdh1}
F_{DH}[\psi({\bf r})]&=&-\frac12\,\varepsilon\varepsilon_0 \int_V\mathrm{d}^3{\bf r}\,\big[\left(\bnabla \psi\right)^2 + \kappa^2 \psi^2\big]\nonumber\\
&+& \oint_{\mathcal{S}}\mathrm{d}S\,\sigma(\Omega)\,\psi(R, \Omega), 
\end{eqnarray}
where $\kappa^2 = 2 e^2 n_{0} /(\varepsilon\varepsilon_0 {k_{\mathrm{B}}T})$ is the square of the inverse Debye screening length. Here, $T$ is the temperature, $k_B$ the Boltzmann constant, and $n_0$ the bulk univalent salt concentration. The surface integral in Eq.~\eqref{eq:fdh1} runs over the entire surface of the sphere. A full PB free energy could also be used here; however, its use would preclude all analytical calculations, which can however illuminate the problem in spite of their approximate nature, as we will show later on.

The second, CR term of the free energy in Eq.~\eqref{eq:fe} can be written as~\cite{Majee2018}
\begin{eqnarray}
F_{CR}[\eta_k]&=&\sum_{k}\alpha_k\eta_k\nonumber\\
&+&\frac{1}{\beta}\sum_{k}\Big[\eta_k\ln\eta_k+(1-\eta_k)\ln(1-\eta_k)\Big],
\end{eqnarray}
and corresponds to the Langmuir-Davies isotherm, or indeed to the Ninham-Parsegian CR condition~\cite{Ninham1971}. Here, $\alpha_k$ represents the non-electrostatic dissociation free energy penalty of the site $k$. Other models of course exist, leading to different surface charging isotherms~\cite{Borkovec2001}.

Next, by minimizing the total free energy [Eq.~\eqref{eq:fe}] with respect to $\psi$, solving for $\psi$, and then evaluating the first two terms in Eq.~\ref{eq:fe}, we remain with
\begin{equation}
\label{eq:fes1}
F=F_{ES}[\sigma(\Omega)]+F_{CR}[\eta_k] = F[\eta_k].
\end{equation}
Here, $F_{ES}[\sigma(\Omega)]$ is now the equilibrium DH electrostatic free energy for a surface charge distribution given by $\sigma(\Omega)$, in {\em absence} of CR. Using furthermore the multipole expansion of the surface charge density, the free energy can be written as
\begin{eqnarray}
\label{eq:fes2}
F_{ES}[\sigma(\Omega)]&=&\frac12\oint_{\mathcal{S}}\mathrm{d}S\,\sigma(\Omega)\,\psi(R, \Omega) \nonumber\\
&=& \frac{1}{8\pi}\oint\mathrm{d}\Omega\,\psi(R, \Omega)\sum_{l,m}\sigma(lm)\,Y_{lm}(\Omega).
\end{eqnarray}
The last equality in Eq.~\eqref{eq:fes2} follows from the definition of the multipole expansion [Eq.~\eqref{eq:mexp}], additionally implying $F_{ES}[\sigma(\Omega)]\rightarrow F_{ES}[\eta_k]$. A full PB electrostatic free energy would differ from the above expression only in the substitution
\begin{equation}
\frac12\,\sigma(\Omega)\,\psi(R, \Omega)\rightarrow\int_{0}^{\sigma(\Omega)}\!\!\!\!\psi(R, \Omega)\,\mathrm{d}\sigma(\Omega),
\end{equation}
corresponding to the Casimir charging process~\cite{Verwey}, with everything else remaining unchanged.

Minimizing now further the expression for the free energy in Eq.~\eqref{eq:fes1} with respect to $\eta_i$, we obtain the CR condition
\begin{equation}
\label{eq:min}
\frac{\partial F_{ES}[\eta_k]}{\partial\eta_i}+\alpha_i+\frac{1}{\beta}\ln\frac{\eta_i}{1-\eta_i }=0.
\end{equation}
In absence of electrostatics ($F_{ES}=0$), the CR condition yields
\begin{equation}
\label{eq:bare}
\eta_i =\frac{1}{1+e^{\beta\alpha_i}},
\end{equation}
where we can identify $\beta\alpha_i=\ln10(pH-pK_a^{(i)})$~\cite{ALB2017a}. Here, $pK_a^{(i)}$ is the association/dissociation constant of the $i$-th charged group. Equation~\eqref{eq:min} then gives the full electrostatic modification of this result:
\begin{equation}
\label{eq:eta}
\eta_i=\left[1+\exp\left(\ln10(pH-pK_a^{(i)})+\beta\frac{\partial F_{ES}[\eta_k]}{\partial\eta_i }\right)\right]^{-1}.
\end{equation}
It is clear by comparison with the results from Ref.~\onlinecite{ALB2017a} that the last term in the exponent in Eq.~\eqref{eq:eta}, $\partial F_{ES}/\partial\eta_i$, plays the role of the local electrostatic potential at site $i$, $\psi_i$. From Eq.~\eqref{eq:eta} we can also immediately obtain the values of individual charges:
\begin{equation}
\label{eq:qkpm}
q_i^\pm=\frac{\pm e}{1+\exp\left(\pm\ln10(pH-pK_a^{(i)})\pm\beta\frac{\partial F_{ES}[\eta_k]}{\partial\eta_i}\right)}.
\end{equation}
To evaluate the derivative of $F_{ES}[\sigma(\Omega)]=F_{ES}[\eta_k]$, we can rewrite it in terms of $\sigma(lm)$, thus obtaining
\begin{equation}
\label{eq:partfes}
\frac{\partial F_{ES}[\sigma(\Omega)]}{\partial\eta_i }=4\pi e\sum_{l,m}\frac{\partial F_{ES}[\sigma(lm)]}{\partial\sigma(lm)}\,g_l(\lambda_i)\,Y_{lm}^*(\Omega_i),
\end{equation}
where we have taken into account the multipole expansion of the surface charge density, Eq.~\eqref{eq:slm}. This expression can be inserted back into Eq.~\eqref{eq:qkpm} to obtain a set of non-linear equations for the charges $q_k$.

In addition, one can also evaluate the equilibrium free energy after the values for the CR charges have been obtained. Inserting Eq.~\eqref{eq:min} into Eq.~\eqref{eq:fes1}, we are left with
\begin{eqnarray}
F[\eta_k]&=&F_{ES}[\eta_k]-\sum_i\frac{\partial F_{ES}[\eta_k]}{\partial\eta_i}\,\eta_i\nonumber\\
&-& \frac{1}{\beta} \sum_i\ln{\left( 1+\exp\left(-\beta\alpha_i-\beta\frac{\partial F_{ES}[\eta_k]}{\partial\eta_i }\right)\right)}. \nonumber\\
~
\end{eqnarray}
In this way, we have expressed the total free energy of the system through its electrostatic part and its derivatives. The above equation is of course valid only for the Langmuir-Davies dissociation isotherm, and different forms would be obtained for different models of CR.

\subsection{Linearized CR approximation}

Since the free energy we have used in our derivation is based on the DH approximation, valid for small electrostatic potentials and thus necessarily for small electrostatic interactions, an expansion of Eq.~\eqref{eq:qkpm} in terms of the electrostatic contribution is in order. This is the logic behind the linear CR approximation previously used to tame the non-linearity of the CR theory~\cite{Carnie1993,Chan1976}, and we use the same argumentation also in our case. We thus proceed with the linear order of the CR condition that leads to the following simplified form:
\begin{equation}
\label{eq:linCR}
q_i=q_i^{(0)}-\beta e\,c_i^{(0)}\frac{\partial F_{ES}[\eta_k]}{\partial\eta_i},
\end{equation}
where $q_i^{(0)}$ is the bare regulated charge of the $i$-th moiety in absence of electrostatics [Eq.~\eqref{eq:bare}]:
\begin{equation}
\label{eq:barechr}
q_i^{(0)}=\frac{\pm e}{1+\exp\left(\pm\ln10(pH-pK_a^{(i)})\right)},
\end{equation}
and $c_i^{(0)}$ is its corresponding (dimensionless) bare capacitance,
\begin{eqnarray}
\label{eq:barecap}
c_i^{(0)}&=&\frac{\exp\left(\ln10(pH-pK_a^{(i)})\right)}{\left[1+\exp\left(\ln10(pH-pK_a^{(i)})\right)\right]^2}\\
&=&\frac{1}{e\ln10}\frac{\partial\,q_i^{(0)}}{\partial\,pH},\nonumber
\end{eqnarray}
The last line explicitly shows that this is indeed a capacitance associated with the $i$-th moiety~\cite{Lund2005}. In this context, we use the term ``bare'' to denote that the electrostatic contribution to these quantities has not been taken into account.

Figure~\ref{fig:1}c shows the $pH$ dependence of the bare charge and bare capacitance of a single positive and negative charge with $pK_a=10$ and $pK_a=4$, respectively. We see that, first of all, when the bare charges are ``fully'' charged, $q_\pm^{(0)}=\pm e$, their bare capacitance is negligible. The latter attains a maximum when the charges reach their mid-point, $q_\pm^{(0)}=\pm e/2$, where  $c_\pm^{(0)}=1/4$. This occurs when $pH$ is equal to the $pK_a$ value of the charge. When the bare charge goes to zero, so does the bare capacitance, with their ratio becoming equal to $1$. With this in mind, we can already predict that CR should have the biggest influence close to the $pK_a$ value of an individual charge, and gradually lose in importance as the $pH$ value moves further away from the $pK_a$.

\subsection{Linearized CR approximation and the multipole expansion}

The linearized form of CR can also be written on the level of the multipole coefficients of the surface charge density $\sigma(lm)$. Inserting Eq.~\eqref{eq:linCR} into Eq.~\eqref{eq:slm}, and at the same time taking into account Eq.~\eqref{eq:partfes}, we are left with
\begin{eqnarray}
\label{eq:slm3}
\sigma(lm)&=&\sigma^{(0)}(lm)-\beta(4\pi e)^2\sum_kc_k^{(0)}\nonumber\\
&\times&\left[\sum_{p,q}\frac{\partial F_{ES}}{\partial\sigma(pq)}\,g_l(\lambda_k)\,g_p(\lambda_k)\,Y_{lm}^*(\Omega_k)\,Y_{pq}^*(\Omega_k)\right],\nonumber\\
~
\end{eqnarray}
which is completely equivalent to the local relation of Eq.~\eqref{eq:linCR}. The expression for $\sigma^{(0)}(lm)$ is identical to the one for $\sigma(lm)$ with $q_k\rightarrow q_k^{(0)}$. In order to stress the physical content of the condition in Eq.~\eqref{eq:slm3}, we can use the well-known formula for the product of two spherical harmonics~\cite{Arfken} and write
\begin{widetext}
\begin{eqnarray}
\label{eq:slm4}
\sigma(lm) = \sigma^{(0)}(lm) - \beta(4\pi e)^2\sum_{L,M}\sum_{p,q} {\cal T}(lm\vert pq\vert LM)\,\frac{\partial F_{ES}[\sigma(pq)]}{\partial\sigma(pq)}\,{\cal C}^{(0)}(LM\vert l\vert p),
\end{eqnarray}
\end{widetext}
where
\begin{eqnarray}
{\cal T}(lm\vert pq\vert LM)&=&\sqrt{\frac{(2l+1)(2p+1)(2L+1)}{4\pi}}\nonumber\\
&\times& \tj{l}{p}{L}{m}{q}{M} \tj{l}{p}{L}{0}{0}{0}
\end{eqnarray}
connects the product of two different spherical harmonic functions to the sum over a single spherical harmonic function, all at the same solid angle. The expressions in parentheses denote the Wigner 3-j symbols. In Eq.~\eqref{eq:slm4} we have also introduced the collective multipole capacitance of the complete charge distribution, obtained by summing over all the sites:
\begin{equation}
{\cal C}^{(0)}(LM\vert l\vert p)=\sum_{k}c_k^{(0)}g_l(\lambda_k)\,g_p(\lambda_k)\,Y_{LM}^*(\Omega_k).
\end{equation}
This expression connects the bare capacitance of the $k$-th site $c_k^{(0)}$ [Eq.~\ref{eq:barecap}] and the measure of the angular size of the $k$-th site $g_l(\lambda_k)$ [Eq.~\ref{eq:gl}].

Equation~\eqref{eq:slm4} thus connects only the collective variables: the multipole moments $\sigma(lm)$ and the multipole capacitance ${\cal C}^{(0)}(LM\vert l\vert p)$. Since the derivative $\partial F_{ES}[\sigma(pq)]/\partial\sigma(pq)$ is a linear function of $\sigma(pq)$ in the DH approximation, it follows that the CR condition mixes different spherical harmonic coefficients in such a way that $M=m+q$ and $\vert\, l-p\vert\leqslant L \leqslant\vert\, l+p\vert$. This admixture of different multipole moments is a simple consequence of the fact that the association/dissociation equilibrium giving rise to CR condition [Eq.~\eqref{eq:qkpm}] self-consistently couples the local charge at each site  with the value of the global electrostatic potential resulting from the charges of all the other sites. The mixing of different multipoles then follows straightforwardly from the connection between the local site variables and the collective multipole moments.

\section{Single sphere with inhomogeneous surface charge distribution}

The derived analytical expressions for the charges and multipole moments in presence of CR can be used in any system where the form of the electrostatic free energy is known. As an example, we will apply them to spherical shells with the same bathing solution on both sides of a thin, proteinaceous surface, such as viral capsids and virus-like particles, as well as to globular proteins or patchy colloids with a dielectric core impermeable to the bathing solution.

In general, the electrostatic free energy of an inhomogeneously charged sphere with radius $R$ can be written in the DH limit as~\cite{ALB2013a,ALB2011}
\begin{equation}
\label{eq:fess}
F_{ES}[\sigma(lm)]=\frac{1}{32\pi^2\varepsilon_w\varepsilon_0R}\sum_lC(l,\kappa R)\sum_m|\sigma(lm)|^2.
\end{equation}
The only difference between the the case of an ion-permeable shell and the case of an impermeable, dielectric sphere lies in the function $C(l,\kappa R)$, which is defined as
\begin{equation}
\label{eq:cp}
C_P(l,x)=I_{l+1/2}(x)\,K_{l+1/2}(x)
\end{equation}
for the ion-permeable shell, and
\begin{equation}
\label{eq:cd}
C_D(l, x)=\frac{1}{x}\left[(\ee-1)\,\frac{l}{x}-\frac{K_{l+3/2}(x)}{K_{l+1/2}(x)}\right]^{-1}
\end{equation}
for the impermeable dielectric sphere. Here, $I_l$ and $K_l$ are the modified Bessel functions of the first and second kind, respectively, $\varepsilon_w$ is the dielectric constant of water, and we have defined for the dielectric sphere the ratio $\ee=\varepsilon_p/\varepsilon_w$, where $\varepsilon_p$ is the dielectric constant of the sphere.

From Eq.~\eqref{eq:fess} we can immediately write down the derivative of the free energy with respect to the chargeable moieties:
\begin{eqnarray}
\frac{\partial F_{ES}[\eta_k]}{\partial\eta_i}&=&\frac{e}{4\pi\varepsilon_w\varepsilon_0R}\sum_{l}(2l+1)\,g_l(\lambda_i)\,C(l,\kappa R)\nonumber\\
&\times&\sum_k q_k\,g_l(\lambda_k)\,P_l(\cos\gamma_{ik}),
\end{eqnarray}
where we have applied the addition theorem for spherical harmonics~\cite{Arfken}. (Alternatively, one could use here also the expression for the derivative of the free energy with respect to the multipole coefficients.) Introducing the Bjerrum length in water, $\ell_B=\beta e^2/4\pi\varepsilon_w\varepsilon_0$, we next define
\begin{equation}
\label{eq:xi}
\xi_{ik}=\frac{\ell_B}{R}\sum_{l}(2l+1)\,g_l(\lambda_i)\,g_l(\lambda_k)\,C(l,\kappa R)\,P_l(\cos\gamma_{ik})
\end{equation}
and thus obtain
\begin{equation}
\beta e\,\frac{\partial F_{ES}[\eta_k]}{\partial\eta_i}=\sum_k\xi_{ik}\,q_k.
\end{equation}
We also note that $\xi_{ik}=\xi_{ki}$. Taking into account the linearized CR approximation [Eq.~\eqref{eq:linCR}], we are left with a linear system of equations:
\begin{equation}
\label{eq:linset}
\sum_k\left(\delta_{ik}+c_i^{(0)}\xi_{ik}\right)q_k=q_i^{(0)}.
\end{equation}
To solve this system of equations, we need the knowledge of the bare charges ($q_i^{(0)}$) and bare capacitances ($c_i^{(0)}$) at a given $pH$, as well as the corresponding coefficients $\xi_{ik}$. From Eq.~\eqref{eq:linset} we can again clearly observe the admixing of different charging sites proportional to the capacitance of a given site.

\subsection{Examples}

Once we obtain the CR-corrected values of charge, we can immediately obtain also the corresponding coefficients of the multipole expansion $\sigma(lm)$, wherefrom we can calculate the multipole magnitudes of the surface charge distribution as~\cite{ALB2018b}
\begin{equation}
S_l=\sqrt{\frac{4\pi}{2l+1}\sum_m\vert\sigma(lm)\vert^2}.
\end{equation}
We will use the magnitudes of the monopole ($\ell=0$), dipole ($\ell=1$), and quadrupole ($\ell=2$) moment to explore the effects of CR on several different spherical surface charge distributions. With $S_l^{(0)}$ we will denote the multipole moments in absence of CR, whereas $S_l$ will be used to refer to multipole moments with CR taken into account. We will keep the temperature of the system fixed ($T=300$ K), while we will mainly vary the salt concentration $n_0$, the radius of the sphere $R$, and the solution $pH$. In addition, we will consider distributions with different numbers of charges $N$, their dissociation constants $pK_a$, and ``patchiness'' $\lambda$ (which will be, for simplicity, assumed identical for all charges in a distribution).

As a first example, we consider a model dipolar patchy distribution on a dielectric, impermeable sphere with $\varepsilon_p=4$. To generate such a distribution, we pick the positions of the chargeable moieties using Mitchell's best candidate algorithm, which randomly places charges on the sphere while preserving some minimal distance between them (for details, see Ref.~\onlinecite{ALB2018b}). Then, to obtain a dipolar distribution, charges on the northern hemisphere ($\vartheta_k\leqslant\pi/2$) are assigned a positive sign ($q_k=+e$), whereas the charges on the southern hemisphere are assigned a negative sign. There are thus $N/2$ positive and $N/2$ negative charges in the distribution. For simplicity, we also assign a $pK_a=10$ to all positive charges and $pK_a=4$ to all negative charges.

\begin{figure}[!tb]
\begin{center}
\includegraphics[width=\columnwidth]{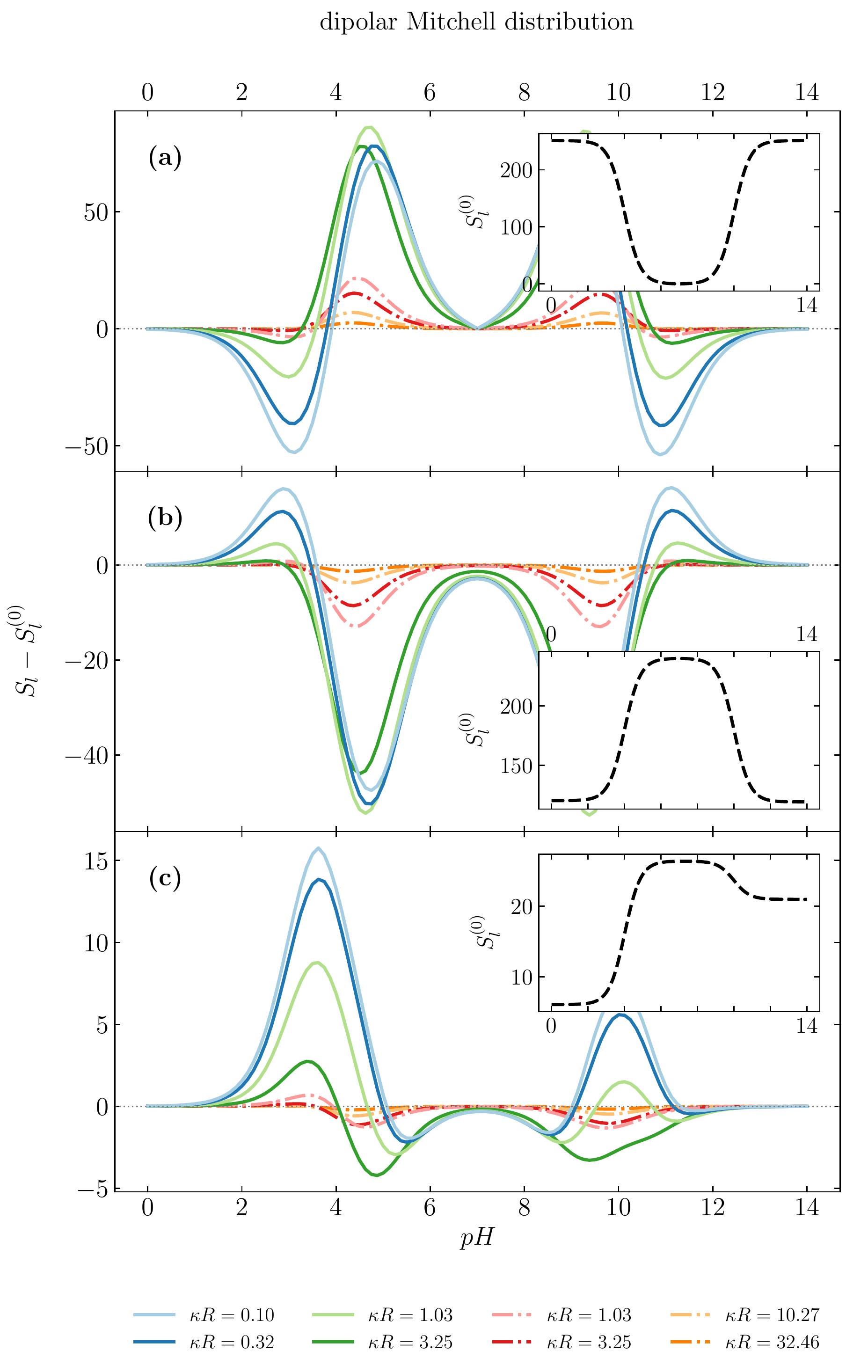}
\end{center}
\caption{Difference between the {\bf (a)} monopole ($\ell=0$), {\bf (b)} dipole ($\ell=1$), and {\bf (c)} quadrupole ($\ell=2$) moments in presence and absence of CR, $S_l-S_l^{(0)}$, as a function of $pH$. The plots are shown for a model patchy dipolar distribution with $N=40$ charges, consisting of $20$ positive charges with $pK_a=10$ and $20$ negative charges with $pK_a=4$ charges, placed on the opposite hemispheres of a sphere using Mitchell's algorithm~\cite{ALB2018b}. Solid curves represent spheres with radius $R=1$ nm and dot-dashed curves spheres with radius $R=10$ nm. Different curves then correspond to different values of the univalent bulk salt concentration ($n_0=1$, $10$, $100$, $1000$ mM). Other parameters of the system are $\lambda=20$ and $\varepsilon_p=4$. To give a sense of the scale of the CR correction, the dashed curves in the insets show the $pH$ dependence of the bare multipole moments $S_l^{(0)}$ in absence of CR.
\label{fig:2}}
\end{figure}

The effects of CR on the first three multipole moments of such a distribution with $N=40$ and $\lambda=20$ are shown in Fig.~\ref{fig:2} for the entire range of $pH$ values, where we plot the difference of the multipole magnitudes in the presence and absence of CR, $S_l-S_l^{(0)}$. To give a sense of the scale of the CR correction, the insets in Fig.~\ref{fig:2} show only the bare multipole magnitudes in absence of CR. We can immediately observe that CR has a significant effect on all of the multipole magnitudes. The effects of CR are particularly large in the vicinity of the $pK_a$ values of the constituent charges, as can be expected from Fig.~\ref{fig:1}c and the fact that the admixing of different higher order multipoles is proportional to the multipole capacitance. Lower salt concentrations, and consequently smaller $\kappa R$, lead to larger effects of CR. As can be inferred from Eq.~\eqref{eq:xi}, the radius and salt concentration have to be treated as separate parameters, and we can observe that the effects of CR become less pronounced when the radius is increased, a change more drastic than when we vary the salt concentration.

\begin{figure}[!tb]
\begin{center}
\includegraphics[width=\columnwidth]{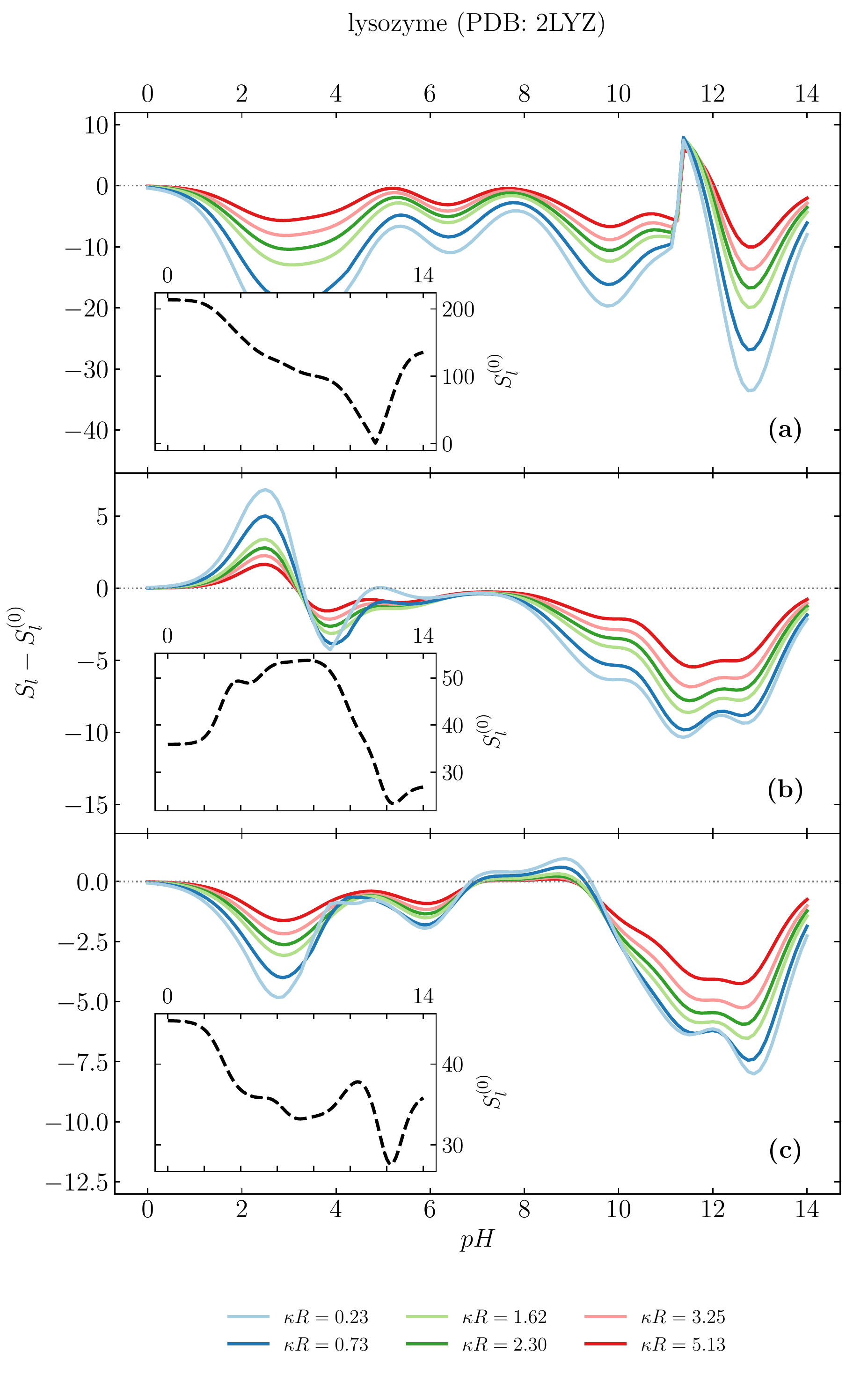}
\end{center}
\caption{Difference between the {\bf (a)} monopole ($\ell=0$), {\bf (b)} dipole ($\ell=1$), and {\bf (c)} quadrupole ($\ell=2$) moments in presence and absence of CR, $S_l-S_l^{(0)}$, as a function of $pH$. The plots are shown for the charge distribution of lysozyme (PDB: 2LYZ), consisting of $28$ positively- and negatively-charged amino acids~\cite{ALB2017a}, whose $pK_a$ values were predicted by PROPKA3.1~\cite{PROPKA3}. The circumscribed radius of the protein is $R=2.24$ nm, and different curves correspond to different values of salt concentration ($n_0=1$, $10$, $50$, $100$, $200$, $500$ mM). Other parameters of the system are $\lambda=20$ and $\varepsilon_p=4$. To give a sense of the scale of the CR correction, the dashed curves in the insets show the $pH$ dependence of the bare multipole moments $S_l^{(0)}$ in absence of CR.
\label{fig:3}}
\end{figure}

\begin{figure}[!tb]
\begin{center}
\includegraphics[width=\columnwidth]{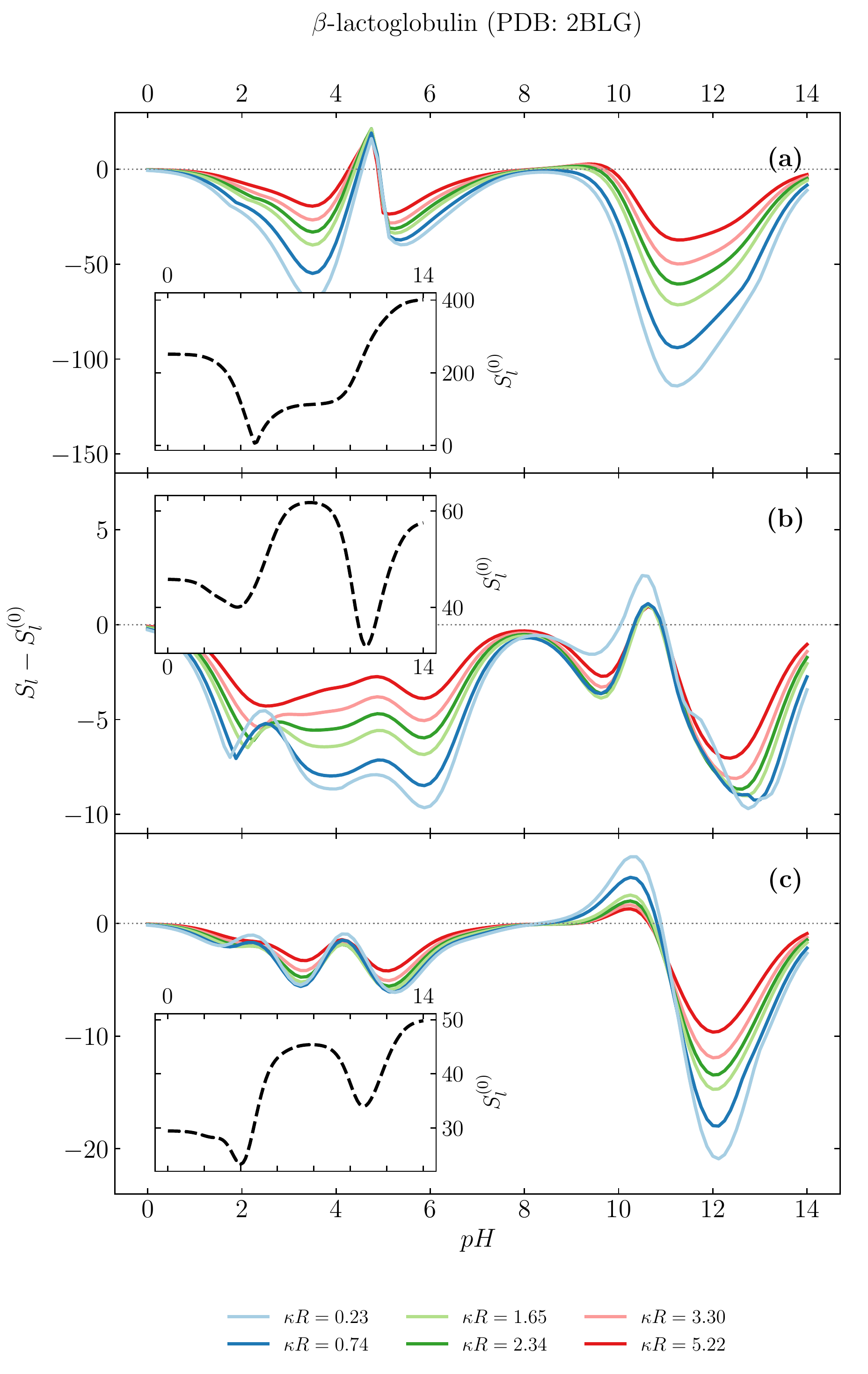}
\end{center}
\caption{Difference between the {\bf (a)} monopole ($\ell=0$), {\bf (b)} dipole ($\ell=1$), and {\bf (c)} quadrupole ($\ell=2$) moments in presence and absence of CR, $S_l-S_l^{(0)}$, as a function of $pH$. The plots are shown for the charge distribution of $\beta$-lactoglobulin (PDB: 2BLG), consisting of $52$ positively- and negatively-charged amino acids~\cite{ALB2017a}, whose $pK_a$ values were predicted by PROPKA3.1~\cite{PROPKA3}. The circumscribed radius of the protein is $R=2.28$ nm, and different curves correspond to different values of salt concentration ($n_0=1$, $10$, $50$, $100$, $200$, $500$ mM). Other parameters of the system are $\lambda=20$ and $\varepsilon_p=4$. To give a sense of the scale of the CR correction, the dashed curves in the insets show the $pH$ dependence of the bare multipole moments $S_l^{(0)}$ in absence of CR.
\label{fig:4}}
\end{figure}

Figure~\ref{fig:2} does not show the influence of other parameters, such as $N$ or $\lambda$, on the effects of CR, as they turn out to be mostly quantitative, influencing to an extent -- but not much -- the magnitude of the CR correction, but not its overall qualitative behaviour. In addition, when the sphere is considered to be permeable to ions, the overall behaviour retains the significance of the CR effects, which could also be expected from the general similarity of the functions in Eqs.~\eqref{eq:cp} and~\eqref{eq:cd}. Lastly, we note that we observe similar behaviour also when constructing model quadrupolar distributions, where the two polar caps are covered with charges of the same type and the equator is covered with the same number of opposite charges.

Next, as an example of more complicated charge distributions, we consider the surface charge distributions of two globular proteins, lysozyme (PDB ID: 2LYZ) and $\beta$-lactoglobulin (PDB ID: 2BLG), whose multipole moments and their $pH$ dependence have already been studied previously in Ref.~\onlinecite{ALB2017a}. The proteins are composed of numerous charged amino acids of different types ($N=28$ for 2LYZ and $N=52$ for 2BLG), whose coordinates were obtained from PDB~\cite{PDB} and whose $pK_a$ values we determined using PROPKA3.1 software~\cite{PROPKA3}. Due to a more complex surface charge distribution, the resulting $pH$ behaviour of the multipole magnitudes becomes more complex than in the previously considered model dipolar and quadrupolar distributions~\cite{ALB2017a}. The two proteins are considered to be impermeable to salt ions with a dielectric constant of $\varepsilon_p=4$, and the patchiness of their constituent charges is taken to be $\lambda=20$ (cf.\ Fig.~\ref{fig:1}b).

The $pH$ dependence of the CR correction to the first three multipole magnitudes is shown in Figs.~\ref{fig:3} and~\ref{fig:4} for 2LYZ and 2BLG, respectively. We see immediately that the scale of CR effects is again significant and comparable to the magnitudes of the bare multipole moments, while their $pH$ dependence is more complicated than in the case of a simple dipolar distribution. We can also observe two rapid shifts in the CR correction of the monopole moments in panels~\ref{fig:3}a and~\ref{fig:4}a, which are related to the isoelectric points of the two proteins (located at approximately $pI\sim11$ and $pI\sim4.5$ for 2LYZ and 2BLG, respectively). As before, the effect of CR is reduced with an increase in salt concentration, yet the CR corrections cannot be neglected even in the limit of high-salt, $n_0\sim1000$ mM. By changing the patchiness of individual charges $\lambda$, the scale of CR corrections is modified, with its effects increasing slightly with an increasing $\lambda$, yet the qualitative behaviour shown in Figs.~\ref{fig:3} and~\ref{fig:4} persists.

The CR effects on the multipole moments of general inhomogeneous charge distributions described in the examples above are thus robust, qualitative, and generally cannot be ignored. Even pronounced electrolyte screening is apparently not enough to completely wipe out the multipole mixing, unless in the highly unlikely case when the screening length would become smaller than the separation between the chargeable moieties on the sphere. This is a simple consequence of the {\em collective effect} of CR, since at each dissociation site the collective electrostatic field of all the other sites contributes to the dissociation reaction.

\section{Conclusions}

We have derived the expressions for the full CR of inhomogeneous spherical surface charge densities and their multipole moments, and solved them explicitly on the level of linear approximation. As a result, we have shown that the CR mechanism leads to a mixing between different multipoles depending on their capacitance (as well as other system parameters), and thus presents an additional anomalous feature of molecular electrostatics. Such effects need to be considered especially in the context of proteins, where multipoles are often assumed to be fixed by structural details.

By considering several examples of inhomogeneously charged spherical particles, we demonstrated that CR can lead to significant changes in the magnitude of their multipole moments even when the system is in the linearized, DH regime. The effects of CR consequently cannot be neglected, and should thus be considered in, e.g., the calculation of the protein-protein interactions, which is often based on multipole expansion. Based on our analysis and the observed effects in terms of multipole mixing, a caveat is also appropriate for computer simulations, where it is obviously quite crucial to incorporate the dissociation equilibrium of the chargeable amino acid moieties in the case of proteins~\cite{Lund2016} or weakly acidic or basic monomers in the case of general polyelectrolytes~\cite{Hofzumahaus2018}.

While the linearized version of the CR condition derived here cannot be readily applied when the electrostatic effects are large, the consistent behaviour of the CR effects upon changes in different system parameters shows that the CR effects should nonetheless persist even in the non-linear electrostatic regime, retaining a considerable effect on the multipole moments. Together with the full directional dependence of the electrostatic interaction in the presence of ionic screening, which involves  involving all multipole moments even in the far-field regime~\cite{Rowan2000,Kjellander2008,ALB2013a,Kjellander2016}, we have therefore identified another anomalous feature of the multipolar expansion in the case of CR in charged macromolecules.

\acknowledgments

ALB and RP acknowledge the financial support from the Slovenian Research Agency (research core funding No.\ (P1-0055)). RP also acknowledges the support of the 1000-Talents Program of the Chinese Foreign Experts Bureau, and the hospitality of the University of the Chinese Academy of Sciences.

\bibliography{references}

\begin{thebibliography}{53}%
\makeatletter
\providecommand \@ifxundefined [1]{%
 \@ifx{#1\undefined}
}%
\providecommand \@ifnum [1]{%
 \ifnum #1\expandafter \@firstoftwo
 \else \expandafter \@secondoftwo
 \fi
}%
\providecommand \@ifx [1]{%
 \ifx #1\expandafter \@firstoftwo
 \else \expandafter \@secondoftwo
 \fi
}%
\providecommand \natexlab [1]{#1}%
\providecommand \enquote  [1]{``#1''}%
\providecommand \bibnamefont  [1]{#1}%
\providecommand \bibfnamefont [1]{#1}%
\providecommand \citenamefont [1]{#1}%
\providecommand \href@noop [0]{\@secondoftwo}%
\providecommand \href [0]{\begingroup \@sanitize@url \@href}%
\providecommand \@href[1]{\@@startlink{#1}\@@href}%
\providecommand \@@href[1]{\endgroup#1\@@endlink}%
\providecommand \@sanitize@url [0]{\catcode `\\12\catcode `\$12\catcode
  `\&12\catcode `\#12\catcode `\^12\catcode `\_12\catcode `\%12\relax}%
\providecommand \@@startlink[1]{}%
\providecommand \@@endlink[0]{}%
\providecommand \url  [0]{\begingroup\@sanitize@url \@url }%
\providecommand \@url [1]{\endgroup\@href {#1}{\urlprefix }}%
\providecommand \urlprefix  [0]{URL }%
\providecommand \Eprint [0]{\href }%
\providecommand \doibase [0]{http://dx.doi.org/}%
\providecommand \selectlanguage [0]{\@gobble}%
\providecommand \bibinfo  [0]{\@secondoftwo}%
\providecommand \bibfield  [0]{\@secondoftwo}%
\providecommand \translation [1]{[#1]}%
\providecommand \BibitemOpen [0]{}%
\providecommand \bibitemStop [0]{}%
\providecommand \bibitemNoStop [0]{.\EOS\space}%
\providecommand \EOS [0]{\spacefactor3000\relax}%
\providecommand \BibitemShut  [1]{\csname bibitem#1\endcsname}%
\let\auto@bib@innerbib\@empty
\bibitem [{\citenamefont {Borkovec}, \citenamefont {J{\"o}nsson},\ and\
  \citenamefont {Koper}(2001)}]{Borkovec2001}%
  \BibitemOpen
  \bibfield  {author} {\bibinfo {author} {\bibfnamefont {M.}~\bibnamefont
  {Borkovec}}, \bibinfo {author} {\bibfnamefont {B.}~\bibnamefont
  {J{\"o}nsson}}, \ and\ \bibinfo {author} {\bibfnamefont {G.~J.}\ \bibnamefont
  {Koper}},\ }in\ \href@noop {} {\emph {\bibinfo {booktitle} {Surface and
  colloid science}}}\ (\bibinfo  {publisher} {Springer},\ \bibinfo {year}
  {2001})\ pp.\ \bibinfo {pages} {99--339}\BibitemShut {NoStop}%
\bibitem [{\citenamefont {{Linderstr{\o}m-Lang}}(1924)}]{Lang1924}%
  \BibitemOpen
  \bibfield  {author} {\bibinfo {author} {\bibfnamefont {K.~U.}\ \bibnamefont
  {{Linderstr{\o}m-Lang}}},\ }\href@noop {} {\bibfield  {journal} {\bibinfo
  {journal} {Medd. Carlsberg Lab.}\ }\textbf {\bibinfo {volume} {15}},\
  \bibinfo {pages} {1} (\bibinfo {year} {1924})}\BibitemShut {NoStop}%
\bibitem [{\citenamefont {Marcus}(1955)}]{Marcus1955}%
  \BibitemOpen
  \bibfield  {author} {\bibinfo {author} {\bibfnamefont {R.}~\bibnamefont
  {Marcus}},\ }\href@noop {} {\bibfield  {journal} {\bibinfo  {journal} {J.
  Chem. Phys.}\ }\textbf {\bibinfo {volume} {23}},\ \bibinfo {pages} {1057}
  (\bibinfo {year} {1955})}\BibitemShut {NoStop}%
\bibitem [{\citenamefont {Tanford}\ and\ \citenamefont
  {Kirkwood}(1957)}]{Tanford1957a}%
  \BibitemOpen
  \bibfield  {author} {\bibinfo {author} {\bibfnamefont {C.}~\bibnamefont
  {Tanford}}\ and\ \bibinfo {author} {\bibfnamefont {J.~G.}\ \bibnamefont
  {Kirkwood}},\ }\href@noop {} {\bibfield  {journal} {\bibinfo  {journal} {J.
  Am. Chem. Soc.}\ }\textbf {\bibinfo {volume} {79}},\ \bibinfo {pages} {5333}
  (\bibinfo {year} {1957})}\BibitemShut {NoStop}%
\bibitem [{\citenamefont {Tanford}(1957)}]{Tanford1957b}%
  \BibitemOpen
  \bibfield  {author} {\bibinfo {author} {\bibfnamefont {C.}~\bibnamefont
  {Tanford}},\ }\href@noop {} {\bibfield  {journal} {\bibinfo  {journal} {J.
  Am. Chem. Soc.}\ }\textbf {\bibinfo {volume} {79}},\ \bibinfo {pages} {5348}
  (\bibinfo {year} {1957})}\BibitemShut {NoStop}%
\bibitem [{\citenamefont {Ninham}\ and\ \citenamefont
  {Parsegian}(1971)}]{Ninham1971}%
  \BibitemOpen
  \bibfield  {author} {\bibinfo {author} {\bibfnamefont {B.~W.}\ \bibnamefont
  {Ninham}}\ and\ \bibinfo {author} {\bibfnamefont {V.~A.}\ \bibnamefont
  {Parsegian}},\ }\href@noop {} {\bibfield  {journal} {\bibinfo  {journal} {J.
  Theor. Biol.}\ }\textbf {\bibinfo {volume} {31}},\ \bibinfo {pages} {405}
  (\bibinfo {year} {1971})}\BibitemShut {NoStop}%
\bibitem [{\citenamefont {Chan}\ \emph {et~al.}(1975)\citenamefont {Chan},
  \citenamefont {Perram}, \citenamefont {White},\ and\ \citenamefont
  {Healy}}]{Chan1975}%
  \BibitemOpen
  \bibfield  {author} {\bibinfo {author} {\bibfnamefont {D.}~\bibnamefont
  {Chan}}, \bibinfo {author} {\bibfnamefont {J.~W.}\ \bibnamefont {Perram}},
  \bibinfo {author} {\bibfnamefont {L.~R.}\ \bibnamefont {White}}, \ and\
  \bibinfo {author} {\bibfnamefont {T.~W.}\ \bibnamefont {Healy}},\ }\href@noop
  {} {\bibfield  {journal} {\bibinfo  {journal} {J. Chem. Soc. Faraday Trans.
  I}\ }\textbf {\bibinfo {volume} {71}},\ \bibinfo {pages} {1046} (\bibinfo
  {year} {1975})}\BibitemShut {NoStop}%
\bibitem [{\citenamefont {Prieve}\ and\ \citenamefont
  {Ruckenstein}(1976)}]{Prieve1976}%
  \BibitemOpen
  \bibfield  {author} {\bibinfo {author} {\bibfnamefont {D.~C.}\ \bibnamefont
  {Prieve}}\ and\ \bibinfo {author} {\bibfnamefont {E.}~\bibnamefont
  {Ruckenstein}},\ }\href@noop {} {\bibfield  {journal} {\bibinfo  {journal}
  {J. Theor. Biol.}\ }\textbf {\bibinfo {volume} {56}},\ \bibinfo {pages} {205}
  (\bibinfo {year} {1976})}\BibitemShut {NoStop}%
\bibitem [{\citenamefont {Pericet-Camara}\ \emph {et~al.}(2004)\citenamefont
  {Pericet-Camara}, \citenamefont {Papastavrou}, \citenamefont {Behrens},\ and\
  \citenamefont {Borkovec}}]{PericetCamara2004}%
  \BibitemOpen
  \bibfield  {author} {\bibinfo {author} {\bibfnamefont {R.}~\bibnamefont
  {Pericet-Camara}}, \bibinfo {author} {\bibfnamefont {G.}~\bibnamefont
  {Papastavrou}}, \bibinfo {author} {\bibfnamefont {S.~H.}\ \bibnamefont
  {Behrens}}, \ and\ \bibinfo {author} {\bibfnamefont {M.}~\bibnamefont
  {Borkovec}},\ }\href@noop {} {\bibfield  {journal} {\bibinfo  {journal} {J.
  Phys. Chem. B}\ }\textbf {\bibinfo {volume} {108}},\ \bibinfo {pages} {19467}
  (\bibinfo {year} {2004})}\BibitemShut {NoStop}%
\bibitem [{\citenamefont {{von Grunberg}}(1999)}]{Grunberg1999}%
  \BibitemOpen
  \bibfield  {author} {\bibinfo {author} {\bibfnamefont {H.~H.}\ \bibnamefont
  {{von Grunberg}}},\ }\href@noop {} {\bibfield  {journal} {\bibinfo  {journal}
  {J. Colloid Interface Sci.}\ }\textbf {\bibinfo {volume} {219}},\ \bibinfo
  {pages} {339} (\bibinfo {year} {1999})}\BibitemShut {NoStop}%
\bibitem [{\citenamefont {Podgornik}\ and\ \citenamefont
  {Parsegian}(1995)}]{Podgornik1995}%
  \BibitemOpen
  \bibfield  {author} {\bibinfo {author} {\bibfnamefont {R.}~\bibnamefont
  {Podgornik}}\ and\ \bibinfo {author} {\bibfnamefont {V.~A.}\ \bibnamefont
  {Parsegian}},\ }\href@noop {} {\bibfield  {journal} {\bibinfo  {journal} {J.
  Chem. Phys.}\ }\textbf {\bibinfo {volume} {99}},\ \bibinfo {pages} {9491}
  (\bibinfo {year} {1995})}\BibitemShut {NoStop}%
\bibitem [{\citenamefont {Henle}\ \emph {et~al.}(2004)\citenamefont {Henle},
  \citenamefont {Santangelo}, \citenamefont {Patel},\ and\ \citenamefont
  {Pincus}}]{Henle2004}%
  \BibitemOpen
  \bibfield  {author} {\bibinfo {author} {\bibfnamefont {M.~L.}\ \bibnamefont
  {Henle}}, \bibinfo {author} {\bibfnamefont {C.~D.}\ \bibnamefont
  {Santangelo}}, \bibinfo {author} {\bibfnamefont {D.~M.}\ \bibnamefont
  {Patel}}, \ and\ \bibinfo {author} {\bibfnamefont {P.~A.}\ \bibnamefont
  {Pincus}},\ }\href@noop {} {\bibfield  {journal} {\bibinfo  {journal}
  {Europhys. Lett.}\ }\textbf {\bibinfo {volume} {66}},\ \bibinfo {pages} {284}
  (\bibinfo {year} {2004})}\BibitemShut {NoStop}%
\bibitem [{\citenamefont {Longo}\ and\ \citenamefont {{Olvera de la Cruz }and
  I.~Szleifer}(2012)}]{Longo2012}%
  \BibitemOpen
  \bibfield  {author} {\bibinfo {author} {\bibfnamefont {G.~S.}\ \bibnamefont
  {Longo}}\ and\ \bibinfo {author} {\bibfnamefont {M.}~\bibnamefont {{Olvera de
  la Cruz }and I.~Szleifer}},\ }\href@noop {} {\bibfield  {journal} {\bibinfo
  {journal} {Soft Matter}\ }\textbf {\bibinfo {volume} {8}},\ \bibinfo {pages}
  {1344} (\bibinfo {year} {2012})}\BibitemShut {NoStop}%
\bibitem [{\citenamefont {Longo}, \citenamefont {{Olvera de la Cruz}},\ and\
  \citenamefont {Szleifer}(2013)}]{Longo2013}%
  \BibitemOpen
  \bibfield  {author} {\bibinfo {author} {\bibfnamefont {G.~S.}\ \bibnamefont
  {Longo}}, \bibinfo {author} {\bibfnamefont {M.}~\bibnamefont {{Olvera de la
  Cruz}}}, \ and\ \bibinfo {author} {\bibfnamefont {I.}~\bibnamefont
  {Szleifer}},\ }\href@noop {} {\bibfield  {journal} {\bibinfo  {journal} {ACS
  Nano}\ }\textbf {\bibinfo {volume} {7}},\ \bibinfo {pages} {2693} (\bibinfo
  {year} {2013})}\BibitemShut {NoStop}%
\bibitem [{\citenamefont {Ad\v{z}i\'{c}}\ and\ \citenamefont
  {Podgornik}(2014)}]{Adzic2014}%
  \BibitemOpen
  \bibfield  {author} {\bibinfo {author} {\bibfnamefont {N.}~\bibnamefont
  {Ad\v{z}i\'{c}}}\ and\ \bibinfo {author} {\bibfnamefont {R.}~\bibnamefont
  {Podgornik}},\ }\href@noop {} {\bibfield  {journal} {\bibinfo  {journal}
  {Eur. Phys. J. E}\ }\textbf {\bibinfo {volume} {37}},\ \bibinfo {pages} {49}
  (\bibinfo {year} {2014})}\BibitemShut {NoStop}%
\bibitem [{\citenamefont {Ad{\v{z}}i{\'c}}\ and\ \citenamefont
  {Podgornik}(2015)}]{Adzic2015}%
  \BibitemOpen
  \bibfield  {author} {\bibinfo {author} {\bibfnamefont {N.}~\bibnamefont
  {Ad{\v{z}}i{\'c}}}\ and\ \bibinfo {author} {\bibfnamefont {R.}~\bibnamefont
  {Podgornik}},\ }\href@noop {} {\bibfield  {journal} {\bibinfo  {journal}
  {Phys. Rev. E}\ }\textbf {\bibinfo {volume} {91}},\ \bibinfo {pages} {022715}
  (\bibinfo {year} {2015})}\BibitemShut {NoStop}%
\bibitem [{\citenamefont {Maggs}\ and\ \citenamefont
  {Podgornik}(2014)}]{Maggs2014}%
  \BibitemOpen
  \bibfield  {author} {\bibinfo {author} {\bibfnamefont {A.~C.}\ \bibnamefont
  {Maggs}}\ and\ \bibinfo {author} {\bibfnamefont {R.}~\bibnamefont
  {Podgornik}},\ }\href@noop {} {\bibfield  {journal} {\bibinfo  {journal}
  {Eur. Phys. J. E}\ }\textbf {\bibinfo {volume} {108}},\ \bibinfo {pages}
  {68003} (\bibinfo {year} {2014})}\BibitemShut {NoStop}%
\bibitem [{\citenamefont {Markovich}, \citenamefont {Andelman},\ and\
  \citenamefont {Podgornik}(2014)}]{Markovich2014}%
  \BibitemOpen
  \bibfield  {author} {\bibinfo {author} {\bibfnamefont {T.}~\bibnamefont
  {Markovich}}, \bibinfo {author} {\bibfnamefont {D.}~\bibnamefont {Andelman}},
  \ and\ \bibinfo {author} {\bibfnamefont {R.}~\bibnamefont {Podgornik}},\
  }\href@noop {} {\bibfield  {journal} {\bibinfo  {journal} {EPL}\ }\textbf
  {\bibinfo {volume} {106}},\ \bibinfo {pages} {16002} (\bibinfo {year}
  {2014})}\BibitemShut {NoStop}%
\bibitem [{\citenamefont {Markovich}, \citenamefont {Andelman},\ and\
  \citenamefont {Podgornik}(2016)}]{Markovich2016}%
  \BibitemOpen
  \bibfield  {author} {\bibinfo {author} {\bibfnamefont {T.}~\bibnamefont
  {Markovich}}, \bibinfo {author} {\bibfnamefont {D.}~\bibnamefont {Andelman}},
  \ and\ \bibinfo {author} {\bibfnamefont {R.}~\bibnamefont {Podgornik}},\
  }\href@noop {} {\bibfield  {journal} {\bibinfo  {journal} {EPL}\ }\textbf
  {\bibinfo {volume} {113}},\ \bibinfo {pages} {26004} (\bibinfo {year}
  {2016})}\BibitemShut {NoStop}%
\bibitem [{\citenamefont {Diamant}\ and\ \citenamefont
  {Andelman}(1996)}]{Diamant1996}%
  \BibitemOpen
  \bibfield  {author} {\bibinfo {author} {\bibfnamefont {H.}~\bibnamefont
  {Diamant}}\ and\ \bibinfo {author} {\bibfnamefont {D.}~\bibnamefont
  {Andelman}},\ }\href@noop {} {\bibfield  {journal} {\bibinfo  {journal} {J.
  Phys. Chem.}\ }\textbf {\bibinfo {volume} {100}},\ \bibinfo {pages} {13732}
  (\bibinfo {year} {1996})}\BibitemShut {NoStop}%
\bibitem [{\citenamefont {Biesheuvel}(2004)}]{Biesheuvel2004}%
  \BibitemOpen
  \bibfield  {author} {\bibinfo {author} {\bibfnamefont {P.~M.}\ \bibnamefont
  {Biesheuvel}},\ }\href@noop {} {\bibfield  {journal} {\bibinfo  {journal} {J.
  Colloid Interface Sci.}\ }\textbf {\bibinfo {volume} {257}},\ \bibinfo
  {pages} {514} (\bibinfo {year} {2004})}\BibitemShut {NoStop}%
\bibitem [{\citenamefont {{Ben-Yaakov}}\ \emph {et~al.}(2011)\citenamefont
  {{Ben-Yaakov}}, \citenamefont {Andelman}, \citenamefont {Podgornik},\ and\
  \citenamefont {Harries}}]{BenYaakov2011}%
  \BibitemOpen
  \bibfield  {author} {\bibinfo {author} {\bibfnamefont {D.}~\bibnamefont
  {{Ben-Yaakov}}}, \bibinfo {author} {\bibfnamefont {D.}~\bibnamefont
  {Andelman}}, \bibinfo {author} {\bibfnamefont {R.}~\bibnamefont {Podgornik}},
  \ and\ \bibinfo {author} {\bibfnamefont {D.}~\bibnamefont {Harries}},\
  }\href@noop {} {\bibfield  {journal} {\bibinfo  {journal} {Curr. Opin.
  Colloid Interface Sci.}\ }\textbf {\bibinfo {volume} {16}},\ \bibinfo {pages}
  {542} (\bibinfo {year} {2011})}\BibitemShut {NoStop}%
\bibitem [{\citenamefont {Trefalt}, \citenamefont {Behrens},\ and\
  \citenamefont {Borkovec}(2016)}]{Trefalt2016}%
  \BibitemOpen
  \bibfield  {author} {\bibinfo {author} {\bibfnamefont {G.}~\bibnamefont
  {Trefalt}}, \bibinfo {author} {\bibfnamefont {S.~H.}\ \bibnamefont
  {Behrens}}, \ and\ \bibinfo {author} {\bibfnamefont {M.}~\bibnamefont
  {Borkovec}},\ }\href@noop {} {\bibfield  {journal} {\bibinfo  {journal}
  {Langmuir}\ }\textbf {\bibinfo {volume} {32}},\ \bibinfo {pages} {380−}
  (\bibinfo {year} {2016})}\BibitemShut {NoStop}%
\bibitem [{\citenamefont {Warshel}\ \emph {et~al.}(2006)\citenamefont
  {Warshel}, \citenamefont {Sharma}, \citenamefont {Kato},\ and\ \citenamefont
  {Parson}}]{Warshel2006}%
  \BibitemOpen
  \bibfield  {author} {\bibinfo {author} {\bibfnamefont {A.}~\bibnamefont
  {Warshel}}, \bibinfo {author} {\bibfnamefont {P.~K.}\ \bibnamefont {Sharma}},
  \bibinfo {author} {\bibfnamefont {M.}~\bibnamefont {Kato}}, \ and\ \bibinfo
  {author} {\bibfnamefont {W.~W.}\ \bibnamefont {Parson}},\ }\href@noop {}
  {\bibfield  {journal} {\bibinfo  {journal} {Biochim. Biophys. Acta}\ }\textbf
  {\bibinfo {volume} {1764}},\ \bibinfo {pages} {1647} (\bibinfo {year}
  {2006})}\BibitemShut {NoStop}%
\bibitem [{\citenamefont {Gitlin}, \citenamefont {Carbeck},\ and\ \citenamefont
  {Whitesides}(2006)}]{Gitlin2006}%
  \BibitemOpen
  \bibfield  {author} {\bibinfo {author} {\bibfnamefont {I.}~\bibnamefont
  {Gitlin}}, \bibinfo {author} {\bibfnamefont {J.~D.}\ \bibnamefont {Carbeck}},
  \ and\ \bibinfo {author} {\bibfnamefont {G.~M.}\ \bibnamefont {Whitesides}},\
  }\href@noop {} {\bibfield  {journal} {\bibinfo  {journal} {Angew. Chem. Int.
  Ed.}\ }\textbf {\bibinfo {volume} {45}},\ \bibinfo {pages} {3022} (\bibinfo
  {year} {2006})}\BibitemShut {NoStop}%
\bibitem [{\citenamefont {Lund}\ and\ \citenamefont
  {J{\"o}nsson}(2005)}]{Lund2005}%
  \BibitemOpen
  \bibfield  {author} {\bibinfo {author} {\bibfnamefont {M.}~\bibnamefont
  {Lund}}\ and\ \bibinfo {author} {\bibfnamefont {B.}~\bibnamefont
  {J{\"o}nsson}},\ }\href@noop {} {\bibfield  {journal} {\bibinfo  {journal}
  {Biochemistry}\ }\textbf {\bibinfo {volume} {44}},\ \bibinfo {pages} {5722}
  (\bibinfo {year} {2005})}\BibitemShut {NoStop}%
\bibitem [{\citenamefont {Krishnan}(2017)}]{Krishnan2017}%
  \BibitemOpen
  \bibfield  {author} {\bibinfo {author} {\bibfnamefont {M.}~\bibnamefont
  {Krishnan}},\ }\href@noop {} {\bibfield  {journal} {\bibinfo  {journal} {J.
  Chem. Phys.}\ }\textbf {\bibinfo {volume} {146}},\ \bibinfo {pages} {205101}
  (\bibinfo {year} {2017})}\BibitemShut {NoStop}%
\bibitem [{\citenamefont {Gramada}\ and\ \citenamefont
  {Bourne}(2006)}]{Gramada2006}%
  \BibitemOpen
  \bibfield  {author} {\bibinfo {author} {\bibfnamefont {A.}~\bibnamefont
  {Gramada}}\ and\ \bibinfo {author} {\bibfnamefont {P.~E.}\ \bibnamefont
  {Bourne}},\ }\href@noop {} {\bibfield  {journal} {\bibinfo  {journal} {BMC
  Bioinformatics}\ }\textbf {\bibinfo {volume} {7}},\ \bibinfo {pages} {242}
  (\bibinfo {year} {2006})}\BibitemShut {NoStop}%
\bibitem [{\citenamefont {Hoppe}(2013)}]{Hoppe2013}%
  \BibitemOpen
  \bibfield  {author} {\bibinfo {author} {\bibfnamefont {T.}~\bibnamefont
  {Hoppe}},\ }\href@noop {} {\bibfield  {journal} {\bibinfo  {journal} {J.
  Chem. Phys.}\ }\textbf {\bibinfo {volume} {138}},\ \bibinfo {pages}
  {05B603\_1} (\bibinfo {year} {2013})}\BibitemShut {NoStop}%
\bibitem [{\citenamefont {{Lo\v{s}dorfer Bo\v{z}i\v{c}}}\ and\ \citenamefont
  {Podgornik}(2017)}]{ALB2017a}%
  \BibitemOpen
  \bibfield  {author} {\bibinfo {author} {\bibfnamefont {A.}~\bibnamefont
  {{Lo\v{s}dorfer Bo\v{z}i\v{c}}}}\ and\ \bibinfo {author} {\bibfnamefont
  {R.}~\bibnamefont {Podgornik}},\ }\href@noop {} {\bibfield  {journal}
  {\bibinfo  {journal} {Biophys. J.}\ }\textbf {\bibinfo {volume} {113}},\
  \bibinfo {pages} {1454} (\bibinfo {year} {2017})}\BibitemShut {NoStop}%
\bibitem [{\citenamefont {Bianchi}\ \emph {et~al.}(2017)\citenamefont
  {Bianchi}, \citenamefont {Capone}, \citenamefont {Coluzza}, \citenamefont
  {Rovigatti},\ and\ \citenamefont {van Oostrum}}]{Bianchi2017}%
  \BibitemOpen
  \bibfield  {author} {\bibinfo {author} {\bibfnamefont {E.}~\bibnamefont
  {Bianchi}}, \bibinfo {author} {\bibfnamefont {B.}~\bibnamefont {Capone}},
  \bibinfo {author} {\bibfnamefont {I.}~\bibnamefont {Coluzza}}, \bibinfo
  {author} {\bibfnamefont {L.}~\bibnamefont {Rovigatti}}, \ and\ \bibinfo
  {author} {\bibfnamefont {P.~D.}\ \bibnamefont {van Oostrum}},\ }\href@noop {}
  {\bibfield  {journal} {\bibinfo  {journal} {Phys. Chem. Chem. Phys.}\
  }\textbf {\bibinfo {volume} {19}},\ \bibinfo {pages} {19847} (\bibinfo {year}
  {2017})}\BibitemShut {NoStop}%
\bibitem [{\citenamefont {Abrikosov}, \citenamefont {Stenqvist},\ and\
  \citenamefont {Lund}(2017)}]{Abrikosov2017}%
  \BibitemOpen
  \bibfield  {author} {\bibinfo {author} {\bibfnamefont {A.~I.}\ \bibnamefont
  {Abrikosov}}, \bibinfo {author} {\bibfnamefont {B.}~\bibnamefont
  {Stenqvist}}, \ and\ \bibinfo {author} {\bibfnamefont {M.}~\bibnamefont
  {Lund}},\ }\href@noop {} {\bibfield  {journal} {\bibinfo  {journal} {Soft
  matter}\ }\textbf {\bibinfo {volume} {13}},\ \bibinfo {pages} {4591}
  (\bibinfo {year} {2017})}\BibitemShut {NoStop}%
\bibitem [{\citenamefont {Boon}\ and\ \citenamefont {van
  Roij}(2011)}]{Boon2011}%
  \BibitemOpen
  \bibfield  {author} {\bibinfo {author} {\bibfnamefont {N.}~\bibnamefont
  {Boon}}\ and\ \bibinfo {author} {\bibfnamefont {R.}~\bibnamefont {van
  Roij}},\ }\href@noop {} {\bibfield  {journal} {\bibinfo  {journal} {J. Chem.
  Phys.}\ }\textbf {\bibinfo {volume} {134}},\ \bibinfo {pages} {054706}
  (\bibinfo {year} {2011})}\BibitemShut {NoStop}%
\bibitem [{\citenamefont {Lund}(2016)}]{Lund2016}%
  \BibitemOpen
  \bibfield  {author} {\bibinfo {author} {\bibfnamefont {M.}~\bibnamefont
  {Lund}},\ }\href@noop {} {\bibfield  {journal} {\bibinfo  {journal} {Colloids
  Surf. B}\ }\textbf {\bibinfo {volume} {137}},\ \bibinfo {pages} {17}
  (\bibinfo {year} {2016})}\BibitemShut {NoStop}%
\bibitem [{\citenamefont {Paulini}, \citenamefont {M{\"u}ller},\ and\
  \citenamefont {Diederich}(2005)}]{Paulini2005}%
  \BibitemOpen
  \bibfield  {author} {\bibinfo {author} {\bibfnamefont {R.}~\bibnamefont
  {Paulini}}, \bibinfo {author} {\bibfnamefont {K.}~\bibnamefont {M{\"u}ller}},
  \ and\ \bibinfo {author} {\bibfnamefont {F.}~\bibnamefont {Diederich}},\
  }\href@noop {} {\bibfield  {journal} {\bibinfo  {journal} {Angew. Chem. Int.
  Ed.}\ }\textbf {\bibinfo {volume} {44}},\ \bibinfo {pages} {1788} (\bibinfo
  {year} {2005})}\BibitemShut {NoStop}%
\bibitem [{\citenamefont {Postarnakevich}\ and\ \citenamefont
  {Singh}(2009)}]{Postarnakevich2009}%
  \BibitemOpen
  \bibfield  {author} {\bibinfo {author} {\bibfnamefont {N.}~\bibnamefont
  {Postarnakevich}}\ and\ \bibinfo {author} {\bibfnamefont {R.}~\bibnamefont
  {Singh}},\ }in\ \href@noop {} {\emph {\bibinfo {booktitle} {Proceedings of
  the 2009 ACM symposium on Applied Computing}}}\ (\bibinfo {organization}
  {ACM},\ \bibinfo {year} {2009})\ pp.\ \bibinfo {pages} {782--787}\BibitemShut
  {NoStop}%
\bibitem [{\citenamefont {Arzensek}, \citenamefont {Kuzman},\ and\
  \citenamefont {Podgornik}(2015)}]{Arzensek2015}%
  \BibitemOpen
  \bibfield  {author} {\bibinfo {author} {\bibfnamefont {D.}~\bibnamefont
  {Arzensek}}, \bibinfo {author} {\bibfnamefont {D.}~\bibnamefont {Kuzman}}, \
  and\ \bibinfo {author} {\bibfnamefont {R.}~\bibnamefont {Podgornik}},\
  }\href@noop {} {\bibfield  {journal} {\bibinfo  {journal} {J. Phys. Chem. B}\
  }\textbf {\bibinfo {volume} {119}},\ \bibinfo {pages} {10375} (\bibinfo
  {year} {2015})}\BibitemShut {NoStop}%
\bibitem [{\citenamefont {{Lo\v{s}dorfer Bo\v{z}i\v{c}}}\ and\ \citenamefont
  {Podgornik}(2018)}]{ALB2018a}%
  \BibitemOpen
  \bibfield  {author} {\bibinfo {author} {\bibfnamefont {A.}~\bibnamefont
  {{Lo\v{s}dorfer Bo\v{z}i\v{c}}}}\ and\ \bibinfo {author} {\bibfnamefont
  {R.}~\bibnamefont {Podgornik}},\ }\href@noop {} {\bibfield  {journal}
  {\bibinfo  {journal} {J. Phys.: Condens. Matter}\ }\textbf {\bibinfo {volume}
  {30}},\ \bibinfo {pages} {024001} (\bibinfo {year} {2018})}\BibitemShut
  {NoStop}%
\bibitem [{\citenamefont {Schwinger}\ \emph {et~al.}(1998)\citenamefont
  {Schwinger}, \citenamefont {DeRaad~Jr}, \citenamefont {Milton},\ and\
  \citenamefont {Tsai}}]{Schwinger}%
  \BibitemOpen
  \bibfield  {author} {\bibinfo {author} {\bibfnamefont {J.}~\bibnamefont
  {Schwinger}}, \bibinfo {author} {\bibfnamefont {L.~L.}\ \bibnamefont
  {DeRaad~Jr}}, \bibinfo {author} {\bibfnamefont {K.}~\bibnamefont {Milton}}, \
  and\ \bibinfo {author} {\bibfnamefont {W.-Y.}\ \bibnamefont {Tsai}},\
  }\href@noop {} {\emph {\bibinfo {title} {Classical electrodynamics}}}\
  (\bibinfo  {publisher} {Westview Press},\ \bibinfo {year} {1998})\BibitemShut
  {NoStop}%
\bibitem [{\citenamefont {Rowan}, \citenamefont {Hansen},\ and\ \citenamefont
  {Trizac}(2000)}]{Rowan2000}%
  \BibitemOpen
  \bibfield  {author} {\bibinfo {author} {\bibfnamefont {D.}~\bibnamefont
  {Rowan}}, \bibinfo {author} {\bibfnamefont {J.-P.}\ \bibnamefont {Hansen}}, \
  and\ \bibinfo {author} {\bibfnamefont {E.}~\bibnamefont {Trizac}},\
  }\href@noop {} {\bibfield  {journal} {\bibinfo  {journal} {Mol. Phys.}\
  }\textbf {\bibinfo {volume} {98}},\ \bibinfo {pages} {1369} (\bibinfo {year}
  {2000})}\BibitemShut {NoStop}%
\bibitem [{\citenamefont {Kjellander}\ and\ \citenamefont
  {Ramirez}(2008)}]{Kjellander2008}%
  \BibitemOpen
  \bibfield  {author} {\bibinfo {author} {\bibfnamefont {R.}~\bibnamefont
  {Kjellander}}\ and\ \bibinfo {author} {\bibfnamefont {R.}~\bibnamefont
  {Ramirez}},\ }\href@noop {} {\bibfield  {journal} {\bibinfo  {journal} {J.
  Phys.: Condens. Matter}\ }\textbf {\bibinfo {volume} {20}},\ \bibinfo {pages}
  {494209} (\bibinfo {year} {2008})}\BibitemShut {NoStop}%
\bibitem [{\citenamefont {{Lo\v{s}dorfer Bo\v{z}i\v{c}}}\ and\ \citenamefont
  {Podgornik}(2013)}]{ALB2013a}%
  \BibitemOpen
  \bibfield  {author} {\bibinfo {author} {\bibfnamefont {A.}~\bibnamefont
  {{Lo\v{s}dorfer Bo\v{z}i\v{c}}}}\ and\ \bibinfo {author} {\bibfnamefont
  {R.}~\bibnamefont {Podgornik}},\ }\href@noop {} {\bibfield  {journal}
  {\bibinfo  {journal} {J. Chem. Phys.}\ }\textbf {\bibinfo {volume} {138}},\
  \bibinfo {pages} {074902} (\bibinfo {year} {2013})}\BibitemShut {NoStop}%
\bibitem [{\citenamefont {Kjellander}(2016)}]{Kjellander2016}%
  \BibitemOpen
  \bibfield  {author} {\bibinfo {author} {\bibfnamefont {R.}~\bibnamefont
  {Kjellander}},\ }\href@noop {} {\bibfield  {journal} {\bibinfo  {journal} {J.
  Chem. Phys.}\ }\textbf {\bibinfo {volume} {145}},\ \bibinfo {pages} {124503}
  (\bibinfo {year} {2016})}\BibitemShut {NoStop}%
\bibitem [{\citenamefont {{Lo\v{s}dorfer Bo\v{z}i\v{c}}}(2018)}]{ALB2018b}%
  \BibitemOpen
  \bibfield  {author} {\bibinfo {author} {\bibfnamefont {A.}~\bibnamefont
  {{Lo\v{s}dorfer Bo\v{z}i\v{c}}}},\ }\href@noop {} {\bibfield  {journal}
  {\bibinfo  {journal} {Soft Matter}\ }\textbf {\bibinfo {volume} {14}},\
  \bibinfo {pages} {1149} (\bibinfo {year} {2018})}\BibitemShut {NoStop}%
\bibitem [{\citenamefont {Majee}, \citenamefont {Bier},\ and\ \citenamefont
  {Podgornik}(2018)}]{Majee2018}%
  \BibitemOpen
  \bibfield  {author} {\bibinfo {author} {\bibfnamefont {A.}~\bibnamefont
  {Majee}}, \bibinfo {author} {\bibfnamefont {M.}~\bibnamefont {Bier}}, \ and\
  \bibinfo {author} {\bibfnamefont {R.}~\bibnamefont {Podgornik}},\ }\href@noop
  {} {\bibfield  {journal} {\bibinfo  {journal} {Soft Matter}\ }\textbf
  {\bibinfo {volume} {14}},\ \bibinfo {pages} {985} (\bibinfo {year}
  {2018})}\BibitemShut {NoStop}%
\bibitem [{\citenamefont {Verwey}, \citenamefont {Overbeek},\ and\
  \citenamefont {Overbeek}(1999)}]{Verwey}%
  \BibitemOpen
  \bibfield  {author} {\bibinfo {author} {\bibfnamefont {E.~J.~W.}\
  \bibnamefont {Verwey}}, \bibinfo {author} {\bibfnamefont {J.~T.~G.}\
  \bibnamefont {Overbeek}}, \ and\ \bibinfo {author} {\bibfnamefont {J.~T.~G.}\
  \bibnamefont {Overbeek}},\ }\href@noop {} {\emph {\bibinfo {title} {Theory of
  the stability of lyophobic colloids}}}\ (\bibinfo  {publisher} {Courier
  Corporation},\ \bibinfo {year} {1999})\BibitemShut {NoStop}%
\bibitem [{\citenamefont {Carnie}\ and\ \citenamefont
  {Chan}(1993)}]{Carnie1993}%
  \BibitemOpen
  \bibfield  {author} {\bibinfo {author} {\bibfnamefont {S.~L.}\ \bibnamefont
  {Carnie}}\ and\ \bibinfo {author} {\bibfnamefont {D.~Y.}\ \bibnamefont
  {Chan}},\ }\href@noop {} {\bibfield  {journal} {\bibinfo  {journal} {J.
  Colloid Interface Sci.}\ }\textbf {\bibinfo {volume} {161}},\ \bibinfo
  {pages} {260} (\bibinfo {year} {1993})}\BibitemShut {NoStop}%
\bibitem [{\citenamefont {Chan}, \citenamefont {Healy},\ and\ \citenamefont
  {White}(1976)}]{Chan1976}%
  \BibitemOpen
  \bibfield  {author} {\bibinfo {author} {\bibfnamefont {D.}~\bibnamefont
  {Chan}}, \bibinfo {author} {\bibfnamefont {T.~W.}\ \bibnamefont {Healy}}, \
  and\ \bibinfo {author} {\bibfnamefont {L.~R.}\ \bibnamefont {White}},\
  }\href@noop {} {\bibfield  {journal} {\bibinfo  {journal} {J. Chem. Soc.
  Faraday Trans.}\ }\textbf {\bibinfo {volume} {72}},\ \bibinfo {pages} {2844}
  (\bibinfo {year} {1976})}\BibitemShut {NoStop}%
\bibitem [{\citenamefont {Arfken}, \citenamefont {Weber},\ and\ \citenamefont
  {Harris}(2011)}]{Arfken}%
  \BibitemOpen
  \bibfield  {author} {\bibinfo {author} {\bibfnamefont {G.~B.}\ \bibnamefont
  {Arfken}}, \bibinfo {author} {\bibfnamefont {H.~J.}\ \bibnamefont {Weber}}, \
  and\ \bibinfo {author} {\bibfnamefont {F.~E.}\ \bibnamefont {Harris}},\
  }\href@noop {} {\emph {\bibinfo {title} {Mathematical methods for
  physicists}}}\ (\bibinfo  {publisher} {Academic press},\ \bibinfo {year}
  {2011})\BibitemShut {NoStop}%
\bibitem [{\citenamefont {{Lo\v{s}dorfer Bo\v{z}i\v{c}}}, \citenamefont
  {\v{S}iber},\ and\ \citenamefont {Podgornik}(2011)}]{ALB2011}%
  \BibitemOpen
  \bibfield  {author} {\bibinfo {author} {\bibfnamefont {A.}~\bibnamefont
  {{Lo\v{s}dorfer Bo\v{z}i\v{c}}}}, \bibinfo {author} {\bibfnamefont
  {A.}~\bibnamefont {\v{S}iber}}, \ and\ \bibinfo {author} {\bibfnamefont
  {R.}~\bibnamefont {Podgornik}},\ }\href@noop {} {\bibfield  {journal}
  {\bibinfo  {journal} {Phys. Rev. E}\ }\textbf {\bibinfo {volume} {83}},\
  \bibinfo {pages} {041916} (\bibinfo {year} {2011})}\BibitemShut {NoStop}%
\bibitem [{\citenamefont {Olsson}\ \emph {et~al.}(2011)\citenamefont {Olsson},
  \citenamefont {S{\o}ndergaard}, \citenamefont {Rostkowski},\ and\
  \citenamefont {Jensen}}]{PROPKA3}%
  \BibitemOpen
  \bibfield  {author} {\bibinfo {author} {\bibfnamefont {M.~H.}\ \bibnamefont
  {Olsson}}, \bibinfo {author} {\bibfnamefont {C.~R.}\ \bibnamefont
  {S{\o}ndergaard}}, \bibinfo {author} {\bibfnamefont {M.}~\bibnamefont
  {Rostkowski}}, \ and\ \bibinfo {author} {\bibfnamefont {J.~H.}\ \bibnamefont
  {Jensen}},\ }\href@noop {} {\bibfield  {journal} {\bibinfo  {journal} {J.
  Chem. Theory Comput.}\ }\textbf {\bibinfo {volume} {7}},\ \bibinfo {pages}
  {525} (\bibinfo {year} {2011})}\BibitemShut {NoStop}%
\bibitem [{\citenamefont {Berman}\ \emph {et~al.}(2000)\citenamefont {Berman},
  \citenamefont {Westbrook}, \citenamefont {Feng}, \citenamefont {Gilliland},
  \citenamefont {Bhat}, \citenamefont {Weissig}, \citenamefont {Shindyalov},\
  and\ \citenamefont {Bourne.}}]{PDB}%
  \BibitemOpen
  \bibfield  {author} {\bibinfo {author} {\bibfnamefont {H.~M.}\ \bibnamefont
  {Berman}}, \bibinfo {author} {\bibfnamefont {J.}~\bibnamefont {Westbrook}},
  \bibinfo {author} {\bibfnamefont {Z.}~\bibnamefont {Feng}}, \bibinfo {author}
  {\bibfnamefont {G.}~\bibnamefont {Gilliland}}, \bibinfo {author}
  {\bibfnamefont {T.~N.}\ \bibnamefont {Bhat}}, \bibinfo {author}
  {\bibfnamefont {H.}~\bibnamefont {Weissig}}, \bibinfo {author} {\bibfnamefont
  {I.~N.}\ \bibnamefont {Shindyalov}}, \ and\ \bibinfo {author} {\bibfnamefont
  {P.~E.}\ \bibnamefont {Bourne.}},\ }\href {www.rcsb.org} {\bibfield
  {journal} {\bibinfo  {journal} {Nucleic Acids Res.}\ }\textbf {\bibinfo
  {volume} {28}},\ \bibinfo {pages} {235} (\bibinfo {year} {2000})}\BibitemShut
  {NoStop}%
\bibitem [{\citenamefont {Hofzumahaus}, \citenamefont {Hebbeker},\ and\
  \citenamefont {Schneider}(2018)}]{Hofzumahaus2018}%
  \BibitemOpen
  \bibfield  {author} {\bibinfo {author} {\bibfnamefont {C.}~\bibnamefont
  {Hofzumahaus}}, \bibinfo {author} {\bibfnamefont {P.}~\bibnamefont
  {Hebbeker}}, \ and\ \bibinfo {author} {\bibfnamefont {S.}~\bibnamefont
  {Schneider}},\ }\href@noop {} {\bibfield  {journal} {\bibinfo  {journal}
  {Soft Matter}\ } (\bibinfo {year} {2018})}\BibitemShut {NoStop}%
\end{thebibliography}%

\end{document}